\documentclass[prf
superscriptaddress,
 amsmath,amssymb,
 aps,
longbibliography
]{revtex4-2}

\usepackage{lipsum}
\usepackage{color}
\usepackage{natbib}
\usepackage{graphicx}
\usepackage{dcolumn}
\usepackage{bm}
\usepackage[caption=false]{subfig}

\begin{document}


\title{Completing Moody's friction diagram in the turbulent transitional regime}

\author{Rory T. Cerbus}
\email{rory.cerbus@riken.jp}
\affiliation{Laboratory for Developmental Epigenetics, RIKEN Center for Biosystems Dynamics Research, Kobe, 650-0047, Japan}
\affiliation{Nonequilibrium Physics of Living Matter RIKEN Hakubi Research Team, RIKEN Center for Biosystems Dynamics Research, Kobe, 650-0047, Japan}

\author{Tom Mullin}
\affiliation{Mathematical Institute, University of Oxford, Oxford, OX26GG, UK}


\date{\today}

\begin{abstract}
The Moody diagram, a plot of friction factor versus flow rate, is a well-known engineering tool for estimating head loss in pipe flows. It comprises well-defined relationships between friction factor and flow rate over the majority of parameter space, but there is a gap in the transitional regime between laminar and turbulent flows. It is often left hatched because in this parameter range the friction is deemed indefinite, which Moody remarked could at least partially be due to the different initial conditions used to establish the flow. Here we investigate this issue and seek a systematic dependence for friction in the transitional regime. The novel method we use is to approach the transitional regime from above by reducing the flow speed from a turbulent flow state. We find that in different pipe flow setups, both driven by gravity, a single curve corresponding to a maximum density of the transitional flow structures is found. We test the generality of this result using an alternative method to drive the flow through the pipe, using a mass displacement device. Our investigation of the flow driven by a syringe produces yet a different curve, indicating that the method of driving the flow has a significant impact on both the final states and the paths to them in the transitional regime.
\end{abstract}

\maketitle

\section{Introduction}
\label{sec:intro}

The Moody diagram is a widely used plot of the relationship between friction, flow speed, and roughness in pipes~\cite{moody1944friction}. It was introduced with the practical goal to ``furnish the engineer with a simple means of estimating the friction factors..."~\cite{moody1944friction}. This was achieved by consolidating and analyzing data gathered from experiments spanning nearly a century into a tool now commonly used by engineers to estimate head loss or practicable pipe diameters~\cite{laviolette2017history}. Although there is room for specific improvement, in particular regarding the treatment of roughness~\cite{flack2018moving}, the essentials are well accepted, and even recent advances are difficult to imagine without Moody in the background. 

The Moody diagram highlights the dependence of friction on flow speed and pipe roughness in a universal way using non-dimensional groups. The non-dimensional friction factor $f$ is defined according to the Darcy formula as $f = (2 D \Delta P/\Delta L)/(\rho U^2)$, where $D$ is the pipe diameter, $\Delta P$ is the pressure drop (or head loss), $\Delta L$ is the pipe length, $\rho$ is the fluid density, and $U$ is the mean flow speed. This is plotted versus the non-dimensional speed or Reynolds number $Re = U D / \nu$, where $\nu$ is the kinematic viscosity. The diagram includes many curves with different non-dimensional roughness $r/D$, where $r$ is the (relative) size of a roughness element~\cite{flack2018moving}. At low $Re$, all the curves conform to the Poiseuille laminar friction formula, $f = f_{\rm{lam}} = 64/Re$. At high $Re$, when the flow is typically turbulent, the roughness becomes important. A lower bound for $f$ is set by a hydraulically smooth pipe which first obeys the turbulent Blasius formula for smooth pipes $f = f_{\rm{turb}} = 0.3164 Re^{-1/4}$ and thereafter the Prandtl friction formula~\cite{swanson2002pipe}. Increasing $r/D$ shifts the curves of friction up and at larger $Re$ it follows the Colebrook equation~\cite{colebrook1939turbulent}. (Other work suggests that $f$ becomes independent of $Re$ at even larger $Re$ to yield the Strickler scaling $f \sim (r/D)^{1/3}$~\cite{gioia2006turbulent}). Friction has been extensively studied in the two extremes of $Re \rightarrow 0$ (laminar) and $Re \rightarrow \infty$ (turbulence), but perhaps in part because of Moody and his diagram, the intermediate region where the flow transitions from laminar to turbulence has been left unresolved. Here we focus on the transitional regime with the immediate objective of searching for friction curves that might fill this gap in the Moody diagram.

\begin{figure}
\centering
\vspace{-0em}
\includegraphics[width=0.9\linewidth]{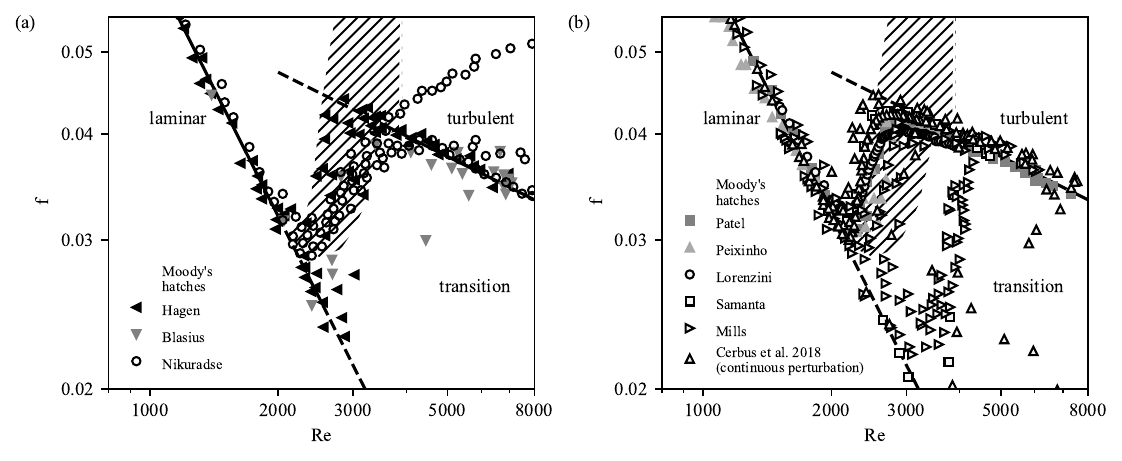}
\vspace{-0.5em}
\caption{Plots of friction factor $f$ vs. Reynolds number $Re$ for a variety of initial conditions. The graphs are constructed using data from a number of different sources~\cite{hagen1855einfluss,blasius1911ahnlickkeitsgesetz,nikuradse1950laws,patel1969some,peixinho2005laminar,lorenzini2010laminar,samanta2011experimental,mills2020high,eckert2021pipe,cerbus2018laws}, including our own previous experiments. The scatter in the data in the transitional region is the key reason why Moody left the critical zone hatched in his original diagram. All experimental data are for Newtonian fluids in smooth pipes apart from the famous experiments of Nikuradse~\cite{nikuradse1950laws}. Examples of well-known experimental data that pre-date Moody. (b) Experimental data that post-date Moody. The Oregon experiments~\cite{swanson2002pipe} were sometimes performed with a screen upstream of the pipe to enhance turbulence. Other experimental data included here were previously produced in the same laboratory with a constant obstacle partially blocking the flow or for a constant injection from the pipe wall~\cite{cerbus2018laws}. By adjusting the amplitude of the perturbation, the size of the obstacle, we may in principle obtain any pair of ($f-Re$) in between the laminar and turbulent curves.}
\label{fig:moodyMayhem}
\vspace{0.0em}
\end{figure}

Moody's original diagram contains a large ``hatched area without definite $f$ lines"~\cite{moody1944friction} connecting the laminar ($f_{\rm{lam}}$) and turbulent ($f_{\rm{turb}}$) curves, as shown in Fig.~\ref{fig:moodyMayhem}(a). Moody remarks that $f$ in this ``critical zone" is characteristically indefinite, with likely reasons being ``the initial turbulence" resulting from imperfections in the pipe, external noise, or pressure waves. He also noted that if the incoming flow has disorder embedded in it, then the flow in the critical zone can even pulsate, as discussed by Prandtl and Tietjens~\cite{tietjens1957applied,cerbus2022prandtl,rotta1956experimenteller}. Reynolds himself searched for friction laws in this transitional regime, but also concluded that the friction here is indefinite~\cite{reynolds1883an}. A fundamental issue for studying the friction in this regime is that the transition from laminar to turbulent flow in pipes is characterized by ostensibly laminar flow with eddying flow interspersed irregularly in time and space~\cite{mullin2011experimental,eckhardt2007turbulence}. At low $Re \lesssim 2250$ this eddying flow is in the form of packets which are $\sim 20D$ long called ``puffs", while at larger $Re \gtrsim 2250$ the packets expand to arbitrary size as they are convected downstream and are eventually termed ``slugs"~\cite{mullin2011experimental,eckhardt2007turbulence,barkley2015rise,avila2023transition}.

Despite this complexity, it was shown that after filtering the friction data to include or exclude eddying flow, the friction factor inside both puffs and slugs follows the turbulent law for smooth pipes ($f_{\rm{turb}}$), and the surrounding laminar flow follows the laminar law ($f_{\rm{lam}}$) for friction~\cite{cerbus2018laws}. Thus the total average of the friction factor over a large pipe section, of the type used to construct the Moody diagram, becomes a function of $Re$ and the flow's turbulent fraction $\gamma$,
\begin{equation}
    f = (1-\gamma) f_{\rm{lam}} + \gamma f_{\rm{turb}} = (1-\gamma) \left[ \frac{64}{Re} \right] + \gamma \left[ 0.3164 Re^{-1/4} \right].
    \label{eq:frictionAverage}
\end{equation}
In this formulation any irregular behavior in $f$ can thus be attributed to $\gamma$. As Moody surmised, the turbulent fraction $\gamma$ certainly depends on the flow's initial conditions, but may also depend on $Re$ and in some cases even $\Delta L$. Indeed many features of the transition between quiescent laminar flow and chaotic turbulent flow depend sensitively on the spatio-temporal details of the triggering perturbation, such as its amplitude or geometry~\cite{hof2003scaling,tasaka2010folded,peixinho2007finite,mellibovsky2006role}. In Fig.~\ref{fig:moodyMayhem}(a) we show a plot of $f$ vs. $Re$, a Moody diagram, in the transitional regime with Moody's hatches using historical data with different experimental conditions which predate Moody \cite{hagen1855einfluss,blasius1911ahnlickkeitsgesetz,eckert2021pipe}. (In this plot and throughout this work we assume the pipes are hydraulically smooth, $r/D \simeq 0$.) The region in between the laminar and turbulent curves, the transitional regime, is indeterminate. In Fig.~\ref{fig:moodyMayhem}(b) we show more recent work~\cite{swanson2002pipe}, including previous experimental data from our own setup which perturbed the flow systematically using an adjustable obstacle~\cite{cerbus2018laws}. The data from Ref.~\cite{swanson2002pipe} can be broadly classified into those with a screen upstream to disturb the incoming flow and those without. With different initial conditions or perturbations, any pair of $(f,Re)$ may be obtained. This casts doubt on the many attempts to find an empirical fit in this region~\cite{joseph2010friction,avci2019new}, and both justifies Moody's decision to leave this region hatched and even suggests that the hatched area should be much larger.


In this work we minimize the influence of initial conditions and determine whether it is possible to establish a single friction curve in the transitional regime. It is well accepted that Poiseuille pipe flow is linearly stable~\cite{kerswell2005} and a finite amplitude disturbance is hence required to cause transition to turbulence~\cite{mullin2011experimental}. In practice the amplitude and form of the disturbance required for transition are both significant factors in the range of $Re$ studied here and obtaining systematic behavior by introducing a disturbance is fraught with difficulties~\cite{darbyshire1995transition}. Hence we attempt to circumvent these potential complications by starting from a well-defined state: turbulence at a value $Re$ above the transition regime. We then perform a ``quench", an experimental protocol wherein we reduce $Re$ to a prescribed value on a very short time scale. We quickly reduce $Re$ to a target value in the transition regime and then observe the ensuing decay of the disordered flow. Relying on the universality of turbulence to study its decay is a well-established technique. Batchelor and Townsend, for example, performed decay experiments starting from the turbulent state to establish that viscous dissipation sets the time scale of the final decay~\cite{batchelor1948decay}. Likewise this quenching approach is similar to previous pipe experiments studying individual puffs~\cite{peixinho2006decay,samanta2011experimental}, transitional plane Couette flow experiments~\cite{bottin1998discontinuous}, and plane Couette-Poiseuille flow experiments~\cite{liu2021decay}.

While our focus is the Moody diagram, we also probe the state of the flow using the time-dependent turbulent fraction $\gamma(t)$, in particular its value after ensemble-averaging over multiple experiments. This has become a common diagnostic for transitional flows, where the regions of turbulent flow are identified by, for example, setting a threshold on the local velocity fluctuations and setting $\gamma(t)$ as the ratio of the size of these regions to the total size of the probed region~\cite{rotta1956experimenteller,mukund2018critical,moxey2010distinct,bottin1998discontinuous}. Here we avoid the ambiguity of a threshold by determining $\gamma(t)$ through the friction factor $f(t)$ by rearranging Eq.~(\ref{eq:frictionAverage}),
\begin{equation}
    \gamma(t) = \frac{f(t) - f_{\rm{lam}}}{f_{\rm{turb}} - f_{\rm{lam}}}.
    \label{eq:gamma}
\end{equation}
Although this method of determining $\gamma(t)$ avoids the inherent ambiguity of setting a threshold on quantities such as the turbulence intensity~\cite{moxey2010distinct}, it also allows for values of $\gamma(t) > 1$ if $f(t) > f_{\rm{turb}}$, which can occur in the early stages of the quench before the flow equilibrates. (While $f_{\rm{lam}}$ is a strict lower bound for $f(t)$, $f_{\rm{turb}}$ is not a strict upper bound~\cite{plasting2005friction}.) 

\section{Constant Pressure Difference (CPD) Experimental Setup}

\begin{figure}
\centering
\vspace{-0em}
\includegraphics[width=0.9\linewidth]{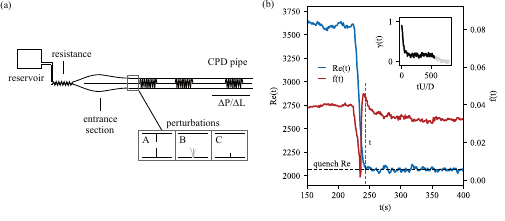}
\vspace{-0.5em}
\caption{(a) Schematic of the Constant Pressure Difference (CPD) pipe experiments modelled after the illustrations in~\citet{reynolds1883an}, with a reservoir and contracting entrance section. Also shown schematically is the resistance external to the experimental pipe section which is employed to minimize flow rate fluctuations. Flow is from left to right. Straight horizontal lines indicate laminar flow, and jagged lines indicates turbulence. The flow may be deliberately perturbed in several ways. The experimental data~\cite{cerbus2018laws} shown in Fig.~\ref{fig:moodyMayhem}(b) used perturbations such as an axially symmetric blockage (A) or injections (B). A small asymmetric obstacle (C) is used to perturb the flow for the quenching experiments and the amount of blockage is varied. (b) Time series of $Re(t)$, $f(t)$ and the resulting $\gamma(t)$ (inset) for a quenching experiment in a $D = 2.5$cm CPD pipe flow. The final quench value is $Re \simeq 2069$. For CPD experiments the $t=0$ point is identified as the time when $Re(t)$ is within the fluctuations about the average final quenched value. The black portion of the $\gamma(t)$ curve is for $tU/D < 560$. After $tU/D \simeq 560$, the undisturbed flow from the entrance has reached the pressure measurement section making $\gamma \rightarrow 0$ ($f \rightarrow f_{\rm{lam}}$).}
\label{fig:setupCPD}
\vspace{0.0em}
\end{figure}

We use two different pipes where in each case the flow is driven by gravity by fixing the height of a reservoir. This sets a constant pressure difference (CPD) along the pipe as can be seen in the schematic diagram of the apparatus in Fig.~\ref{fig:setupCPD}(a). The two CPD flows are similar to other CPD transition pipe experiments~\cite{mullin2011experimental,barkley2015rise}. In our experiments the pipes are each 20-m-long, and constructed using 1-m-long cylindrical glass tubes with inner diameter $D$ = 2.5 cm $\pm$ 10 $\mu$m ($L_{\rm{tot}}/D = 808$), and $D$ = 1 cm $\pm$ 10 $\mu$m  ($L_{\rm{tot}}/D = 2020$), joined by acrylic connectors. 
The CPD flow when disturbance free can remain laminar till the highest $Re$ tested, $Re \approx 10,000$, and we use an external resistance to damp fluctuations in $Re$ so that it remains constant to within $\lesssim$ 1\% and $\lesssim$ 2\% for the $D$ = 2.5 cm and 1 cm pipes, respectively. This has become a standard technique which was first introduced by Rotta to minimize the flow fluctuations in the transitional regime induced by the varying pressure drop in the pipe~\cite{rotta1956experimenteller}. Many modern experimental setups use variants of Rotta's external resistance to reduce fluctuations~\cite{samanta2011experimental,de2009experimental,barkley2015rise,hof2008repeller}. As a result of the pulsations mentioned by Moody, in practice it becomes increasingly difficult to fix $Re$ absolutely in the range of $Re \gtrsim 2250$ where slugs prevail if the flow is driven by a CPD without a very large external resistance~\cite{tietjens1957applied,cerbus2022prandtl}. 


In order to establish a turbulent initial state, we trigger the flow upstream using an asymmetric obstacle. A 10mm-thick Teflon block with a circular hole whose diameter matches the inner diameter of the pipe is housed in a waterproof casing. This block may be moved vertically relative to the pipe so that it can nearly block the entire flow or it can be positioned flush with the wall. We take advantage of the finite amplitude instability of pipe flow~\cite{mullin2011experimental} to adjust the obstacle such that it triggers turbulence only when $Re \gtrsim 3000$, beyond the hatched region in Fig.~\ref{fig:moodyMayhem} where it is possible to establish fluctuations throughout ($\gamma = 1$). In this way we can raise the $Re$ of the flow above $Re \simeq 3000$ to establish a turbulent state. We determined in a series of experiments that details of the quench protocol, such as the initial $Re$ or quench time, are not important. See the Supplementary Material (SM) for details. When we reduce the $Re$ to a  value inside the target transitional regime ($1500 \lesssim Re \lesssim 2700$), the obstacle does not trigger turbulence. We do not consider the flow beyond $t > (L-\Delta L)/U$, the time it takes the fluid from the entrance to reach the pressure measurement section a distance $L-\Delta L$ downstream (see Fig.~\ref{fig:setupCPD}). We use Labview and a DAQ board with all setups to determine $Re(t)$, $f(t)$, and thus $\gamma(t)$, simultaneously. Due to the different time scales of the experiments and limitations of the sensors, the sampling rate for the CPD pipe flows was typically 1Hz. The pressure measurement section of length $\Delta L$ is downstream of the entrance. The temperature remained constant to within $\lesssim 0.03$K during an experiment, which yields variations around the $Re$ of interest of $\delta Re \lesssim 2$. Using two CPD pipes of different diameter allows us to test both the role of pipe length, $L/D$, and quench duration $\Delta T U/ D$, in otherwise identical experimental conditions. For the $D = 2.5$ and $D = 1$ cm pipes, 
$\Delta T U/ D \simeq 33$ and $\Delta T U/ D \simeq 205$, respectively, 
at $Re = 2000$. In App.~\ref{sec:appA} we demonstrate that the final state is insensitive to the details of the quench. 

The experimental protocol is illustrated using the time series data presented in Fig.~\ref{fig:setupCPD}(b). For $t < 0$, the flow is at a high $Re$ and is in a turbulent state until a quench is performed. After this $Re$ is quickly reduced by lowering the reservoir in the CPD experiments, which takes $\simeq$ 11 s ($\Delta T U/ D \simeq 33$ and $\Delta T U/ D \simeq 205$ for the $D = 2.5$ cm and $D = 1$ cm pipes, respectively). 
We define $t = 0$ as the point when the $Re$ is at the target value inside the transitional regime. We use the value of $f(t)$ in the range $0 \leq t \leq L/U$ to determine $\gamma(t)$. We perform each experiment between three and ten times and then ensemble-average the $\gamma(t)$ curves from each run. A typical experiment at one $Re$, which consists of several quench cycles, lasts $\gtrsim 1000$s. 

For some of the experiments an LDV (MSE) was also used to measure the axial velocity time series $u(t)$. The LDV was stationed at a distance of $\sim$ 465$D$ from the entrance of the pipe. Here the pipe was housed inside a rectangular acrylic encasement filled with water to reduce optical distortion as in Ref.~\cite{cerbus2020small}. To conduct LDV measurements, we seeded the water in the pipe with 10$\mu$m density-matched particles. The typical data rate was $>$ 100 Hz.

\section{CPD Results}

\begin{figure}
\centering
\vspace{-0em}
\includegraphics[width=0.9\linewidth]{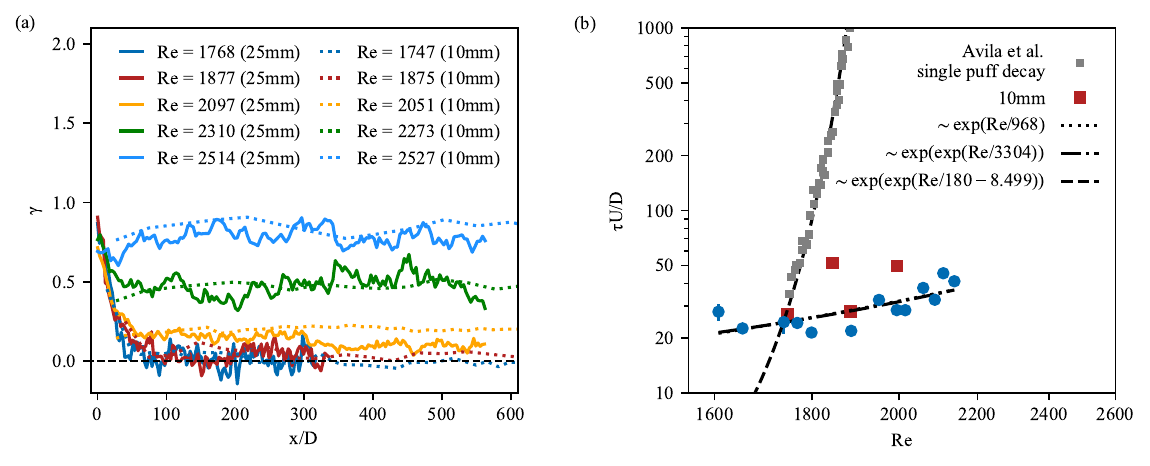}
\vspace{-0.5em}
\caption{(a) Example time series from the two CPD experimental setups at similar $Re$ showing broad agreement with regard to the turbulent fraction $\gamma$. The sampling time for the $D = 10$ mm pipe experiments was in general too slow to determine the earliest stages of the decay. (b) The decay times $\tau U/D$ vs. $Re$ estimated from an exponential fit. The fitting range was until $\gamma$ had reached 0.5 of $(\gamma(t=0) - \gamma(t\rightarrow \infty)$. The uncertainty was determined by fitting to 0.45 and 0.55. Also included are the decay times from CPD experiments studying single puffs~\cite{avila2011onset}. The stark contrast between the scaling of the decay after a quench and for single puffs points to strong interactions between puffs and suggests that these interactions tend to enhance the decay. We also include exponential and super-exponential fits for our experimental data. For this range of data these fits are indistinguishable.
}
\label{fig:CPDTimeSeriesDecayTimes}
\vspace{0.0em}
\end{figure}

We follow the quench protocol for both CPD pipe setups over a range of $Re$ in the turbulent transition regime ($1500 \lesssim Re \lesssim 2700$). In Fig.~\ref{fig:CPDTimeSeriesDecayTimes}(a) we plot several example time series of the turbulent fraction $\gamma(t)$ from both pipe experiments. After an initial exponential decay, the turbulent fraction settles and fluctuates about a steady value. We use the friction factor in this final state, typically averaged over the last 1/5 of the time series, to complete the Moody diagram. The initial decay is difficult to resolve in many of the 10mm-pipe experiments as a result of the relatively poor time resolution, but the averaged final state values are in good agreement between the two CPD pipe experiments. In Fig.~\ref{fig:CPDTimeSeriesDecayTimes}(b) we plot the decay times $\tau$ determined by an exponential fit to the time series before the final state. A moderate increase with $Re$ is found, which is significantly less than those reported for individual puffs~\cite{avila2011onset}. Because the initial state, here turbulence at $Re \gtrsim 3000$, differs from the fluctuating flow inside an individual puff at $Re \lesssim 2250$ a difference between these two curves is not surprising. On the other hand, this also suggests that interactions between turbulent puffs may enhance decay.

\begin{figure}
\centering
\vspace{-0em}
\includegraphics[width=0.9\linewidth]{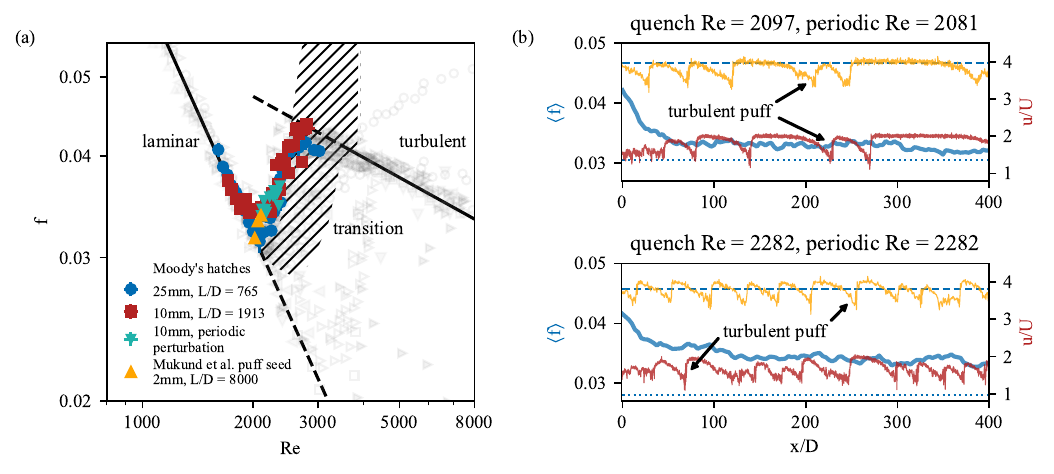}
\vspace{-0.5em}
\caption{(a) Plot of $f$ vs. $Re$ from our quenching experiments, periodic perturbations with maximum turbulent fraction, and long-time experiments starting from a single puff~\cite{mukund2018critical}. The previous scattered data in Fig.~\ref{fig:moodyMayhem}(a) are shown in the background with high transparency for comparison. The results from the different experiments reported here show collapse onto a single curve. (b) Examples of ensemble-averaged friction factor time series with the corresponding velocity time series measured upstream of the pressure measurement section. The velocity was measured with an LDV. The ($---$) and ($\cdot \cdot \cdot$) horizontal lines correspond to the turbulent and laminar friction factors respectively. Also plotted is a periodic perturbation experiment with a similar $Re$ shifted vertically. Both time series show a similar density of puffs.
}
\label{fig:CPDResults}
\vspace{0.0em}
\end{figure}

Next we consider the value of the friction factor in the final state as a function of $Re$, where we have averaged over the end (typically $1/5$) of the ensemble-averaged times series. We plot the combined results of over 100 experiments in Fig.~\ref{fig:CPDResults}(a), where we find that the $f-Re$ data from the quench experiments using two different pipes collapse onto a common curve. While our stated goal is to avoid the complexity of initial conditions, we note that turbulence itself is an amalgamation of different solutions to the Navier-Stokes equation~\cite{willis2016symmetry}, with the number thereof increasing with $Re$~\cite{eckhardt2007turbulence}. It is remarkable that different realizations of turbulence nevertheless yield a regular quenched state, at least in terms of average friction. 

Compared to much of the scattered data in Fig.~\ref{fig:moodyMayhem}, the quenched results have a higher $f$ (higher $\gamma$) at the same $Re$, suggesting that the quench protocol yields a state of maximal $\gamma$. To test this, we performed ancillary experiments whereby we periodically injected 0.15mL of water at 30mL/s to induce puffs, similar to Refs.~\cite{cerbus2018laws,samanta2011experimental}. We varied the period of the injections to achieve a maximum number of puffs, and thus maximal $\gamma$, and measured $f$ in the same way as the quench experiments (see SM). These maximal $f$ relationships also collapse onto the same curve, demonstrating that quenching achieves a maximal $f$ ($\gamma$) through a maximal packing of puffs at a given $Re$. We further probe the role of puffs and their spacing by comparing the velocity time series of the quenching and periodic perturbation experiments in Fig.~\ref{fig:CPDResults}(b). At a similar $Re$, we find that the number of puffs, and thus also their average spacing, is similar for both protocols. Moreover we find that if we estimate $f$ using the number of puffs detected by velocimetry after a quench, their average length~\cite{cerbus2018laws}, and the theoretical value of $f_{\rm{turb}}$, then the estimated $f$ is very close to the value determined by pressure measurements (see Sec.~\ref{sec:comparison}).

Finally, we also include $f$ determined from the reported $\gamma$ from several CPD experiments in a very long pipe seeded with a single puff~\cite{mukund2018critical}. The spatial distribution of puffs at the exit of the pipe was re-created at the entrance using injections to simulate periodic boundary conditions and artificially extend the effective time span of the experiment. Despite the very different initial conditions, these experiments also conform to the same curve, suggesting that this curve is both maximal and stable. Although it can take as little as $\simeq 50$ convection units if turbulent flow is used as an initial condition or as much as $\simeq 10^7$ convection units if the starting condition is a single puff~\cite{mukund2018critical}, our combined data indicate that given enough time most initial conditions should reach this state. We thus conclude that we have established the $f-Re$ curve in the transitional regime and completed the Moody diagram.

\section{Constant Mass Flux (CMF) Experimental Setup}

\begin{figure}
\centering
\vspace{-0em}
\includegraphics[width=0.9\linewidth]{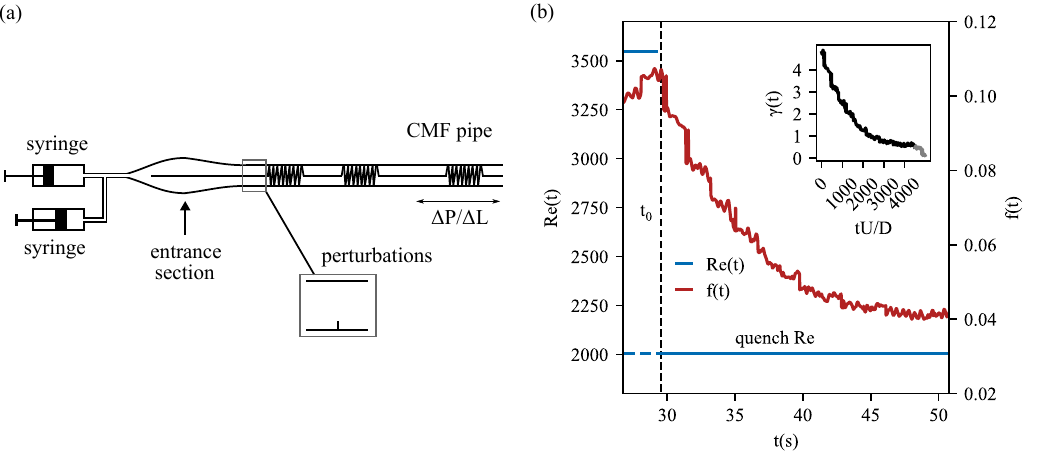}
\vspace{-0.5em}
\caption{(a) Schematic of the Constant Mass Flow (CMF) pipe experiments modelled after the illustrations appearing in~\citet{reynolds1883an}, with two syringes and a contracting entrance section. Flow is from left to right. Straight horizontal lines indicate laminar flow, and jagged lines indicates turbulence. Here we use a small asymmetric blockage to perturb the flow for the quenching experiments and the amount of blockage is varied. (b) Time series of $Re(t)$, $f(t)$ and the resulting $\gamma(t)$ (inset) for a quenching experiment in a $D = 0.3$cm CMF pipe flow. The final quench value is $Re \simeq 2005$. The $t=0$ point is identified as the time when the second syringe has completely ceased moving ($\sim$0.4 s after initiation). The black portion of the $\gamma(t)$ curve is for $tU/D < 4502$. After $tU/D \simeq 4502$, the undisturbed flow from the entrance has reached the pressure measurement section making $\gamma \rightarrow 0$.}
\label{fig:setupCMF}
\vspace{0.0em}
\end{figure}

\begin{figure}
\centering
\vspace{-0em}
\includegraphics[width=0.55\linewidth]{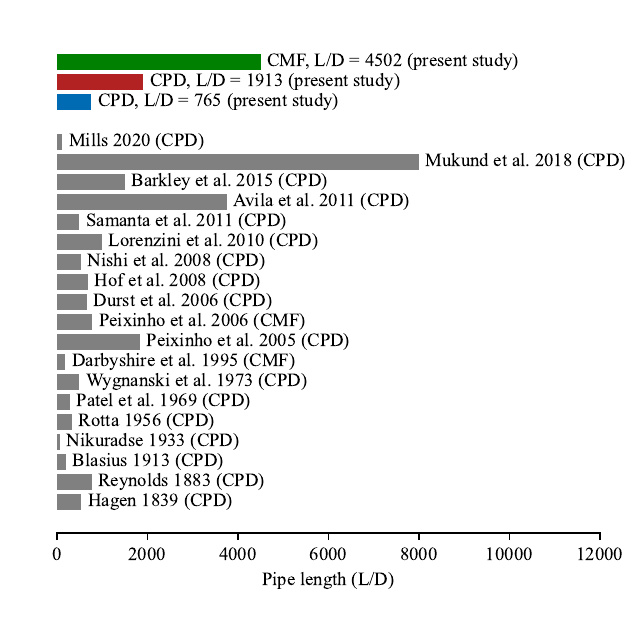}
\vspace{-0.5em}
\caption{Plot of non-dimensional experimental pipe lengths from the studies focusing on the transitional regime~\cite{darbyshire1995transition,rotta1956experimenteller,wygnanski1973transition,nishi2008laminar,durst2006forced,hof2008repeller,peixinho2006decay,barkley2015rise,avila2011onset,mukund2018critical,nikuradse1950laws,lorenzini2010laminar,peixinho2005laminar,hagen1839ueber,blasius1911ahnlickkeitsgesetz,samanta2011experimental,mills2020high} as well as the pipe lengths used in the present study. (If a study included several pipe lengths we chose the maximum length.) The CPD pipe lengths used here are comparable to most previous studies while the CMF pipe length exceeds most of them. We note that the lengths used for previous studies are typically their total pipe length, whereas for our pipes we have included the length only up until the end of the pressure measurement section.}
\label{fig:pipeLengths}
\vspace{0.0em}
\end{figure}

To further test the generality of the friction curve shown in Fig.~\ref{fig:CPDResults}(a), we constructed another pipe experiment which is longer than both CPD pipe experiments. Now, however, the same driving is used as in Refs.~\cite{darbyshire1995transition,peixinho2006decay,peixinho2007finite}, although the pipe used here is longer than these previous experiments. In this driving method a piston acts as a mass displacement device to move the flow~\cite{darbyshire1995transition,peixinho2006decay,swanson2002pipe,hof2003scaling,peixinho2007finite}. This method controls $Re$ directly by driving the flow with a constant mass flux (CMF). Clearly this method can only operate for a finite period of time but when suitably designed this is not a severe limitation in practice. While this method of driving avoids flow rate sensitivity to friction fluctuations, it can potentially introduce mechanical noise which we have nevertheless not observed. It is known that fluctuations in $Re$ are small for both CPD with a large external resistance and effectively absent in CMF flows and hence it might be anticipated that closely similar results will be obtained using either method. However, experiments using the two different driving mechanisms, CPD and CMF, have produced observations of different $Re$-dependence for the decay lifetimes of puffs~\cite{avila2011onset,kuik2010quantitative,peixinho2006decay}, and no puff splitting has been reported in experimental CMF flows~\cite{darbyshire1995transition}. Just as differences between CMF and CPD flow have been noted in a study of subcritical instability of channel flow~\cite{rozhdestvensky1984secondary}, these two sets of experiments suggest that the different driving mechanisms may produce qualitatively distinct outcomes for the decay. However, it has also been argued that any disagreement between the results reported to date can be explained by differences in the lengths of the pipe used in the experiments~\cite{mukund2018critical} or initial conditions~\cite{avila2010transient}. We circumvent these issues by employing a shared quench protocol to establish turbulence as the initial flow condition and by using a long pipe ($L \gtrsim 4000D$) that surpasses the length of most experimental pipe experiments (see Fig.~\ref{fig:pipeLengths}). While previous results suggest that the decay from turbulence may differ between our experiments~\cite{peixinho2006decay,avila2011onset}, there are no previous results which indicate that friction in the steady-state should be dissimilar.

The CMF experimental setup shown in Fig.~\ref{fig:setupCMF}(a) is driven by two independent, computer-controlled, high-pressure syringe pumps (Chemyx Fusion 6000) used in precision microfluidic studies~\cite{gizzatov2021high}. The pipe is 14-m-long, made of 1-m-long sections of cylindrical glass tubes (Duran) of inner diameter $D$ = 0.3 cm $\pm$ 10 $\mu$m ($L_{\rm{tot}}/D = 4840$), joined by 3D-printed and machined plastic connectors. All the connectors have diametrically-opposite holes (of diameter $\sim$ 1 mm) to enable the measurement of the pressure drop $\Delta P(t)$ along the pipe. In the present investigation we focus on measurements near the downstream end of the pipe. 
For the CMF setup the flow speed $U(t)$ is controlled directly 
and we confirmed its accuracy by weighing the amount of water exiting the pipe in a set time period which was measured using a stopwatch. 
The syringe pumps agreed to within $\lesssim 0.1\%$. The flow was conditioned before entering the test section so that it can remain laminar till the highest $Re$ achievable with a single pump, $Re \approx 3,100$. We use Labview and a DAQ board to determine $f(t)$ and thus $\gamma(t)$. The sampling rate for the CMF flow is 100-300Hz. The pressure measurement section of length $\Delta L$ is downstream of the entrance. The temperature variations in the CMF experiments were typically $\lesssim 0.02$ K. 

The quench protocol for the CMF experiments was designed to closely match the protocol in the CPD experiments in several important ways. First, the quench time for the CMF experiments ($\Delta T U/ D \simeq 82$) is intermediate between the two CPD pipe quench times ($\Delta T U /D \simeq $ 33, 205), so that we may safely disregard differences in the quenching time. Likewise the obstacle in each experiment is a scaled version of the obstacle in the other experiments, an asymmetric Teflon obstacle whose size we adjust to only trigger turbulence for $Re \gtrsim 3000$. In all experiments the quench takes place after the fluctuations from the pipe entrance have reached the end of the pipe and we measure until the laminar flow from the entrance finally reaches the pressure measurement section. The CMF quench protocol differs from the CPD quench protocol in the manipulation of the driving necessary to effect the quench. At the beginning of each CMF run we set the first syringe pump at the target flow rate ($Re$) and the second syringe pump such that the combined flow rate yields $Re \gtrsim 3000$ so that the obstacle triggers turbulence. After the entire pipe is turbulent, we stop the second syringe pump so that $Re$ reduces to the target value. The stopping time for the syringe is $\Delta T \sim 0.4$s, which yields the quenching time of $\Delta T U/ D \simeq 82$ at $Re = 2000$. We perform each experiment between three and ten times and then ensemble-average the $\gamma(t)$ curves from each run.

\section{CMF Results}

\begin{figure}
\centering
\vspace{-0em}
\includegraphics[width=1.0\linewidth]{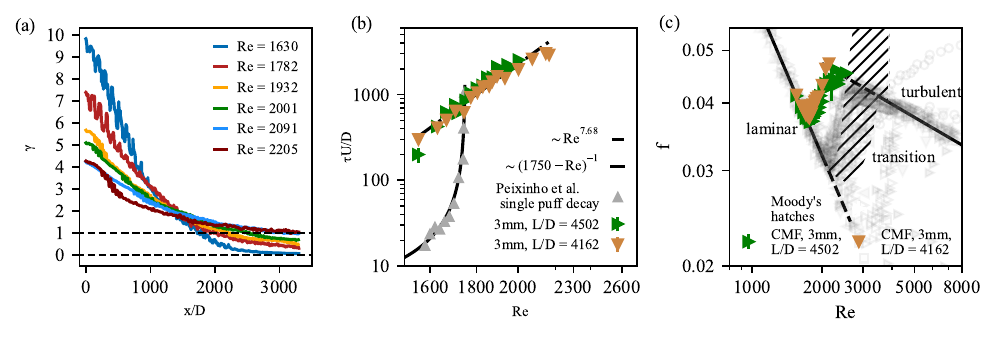}
\vspace{-1.5em}
\caption{(a) Example time series of $\gamma$ from the CMF experimental setup. After an initial steep descent the decay slows to a final state. (b) The decay times $\tau U/D$ vs. $Re$ estimated from an exponential fit. The fitting range was until $\gamma$ had reached 0.5 of $(\gamma(t=0) - \gamma(t\rightarrow \infty)$. The uncertainty was determined by fitting to 0.45 and 0.55. Measurements at two different downstream distances simulates the effect of a shorter pipe. Also included are the decay times from CMF experiments studying single puffs~\cite{peixinho2006decay}. The stark contrast between the scaling of the decay after a quench and for single puffs points to strong interactions between puffs and suggests that in CMF flow these interactions tend to inhibit the decay. (c) Plot of $f$ vs. $Re$ from our CMF quenching experiments using measurements at two different downstream distances to probe the effects of pipe length. We find good agreement between the two. The previous scattered data in Fig.~\ref{fig:moodyMayhem}(a) are shown in the background with high transparency for comparison. The friction curve determined from the CMF experiments lies well above the previous data and Moody's hatches.}
\label{fig:timeSeriesAndDecayAndMoodyCMF}
\vspace{0.0em}
\end{figure}

In Fig.~\ref{fig:timeSeriesAndDecayAndMoodyCMF}(a) we show example time series of the turbulent fraction $\gamma(t)$ for the CMF pipe flow experiments. Just as with the CPD experiments (Fig.~\ref{fig:CPDTimeSeriesDecayTimes}(a)), we find an initial decay followed by a final state. We estimated the decay times as with the CPD experiments and again found a general increase with $Re$, as shown in Fig.~\ref{fig:timeSeriesAndDecayAndMoodyCMF}(b), although as we discuss later the time scales here are significantly longer than in the CPD cases. When compared with decay times reported from studies on single puffs in CMF experiments~\cite{peixinho2006decay}, our decay times tend to be larger. With the caveat that the initial conditions are not identical, this suggests that in CMF flow interactions between puffs may inhibit decay.

Finally, in Fig.~\ref{fig:timeSeriesAndDecayAndMoodyCMF}(c) we show the average value of the friction factor in the final state in a plot of $f$ vs. $Re$, where we have averaged over the end (typically $1/5$) of the ensemble-averaged times series. We use measurements at two different downstream distances to simulate different pipe lengths and find good agreement, although there is a slight tendency for the shorter pipe to have a higher friction. In the transition regime the common friction curve lies above nearly all previously measured $f-Re$ values and also above Moody's hatches. 




\section{Comparison of CMF and CPD experimental results}
\label{sec:comparison}


\begin{figure}
\centering
\vspace{-0em}
\includegraphics[width=0.9\linewidth]{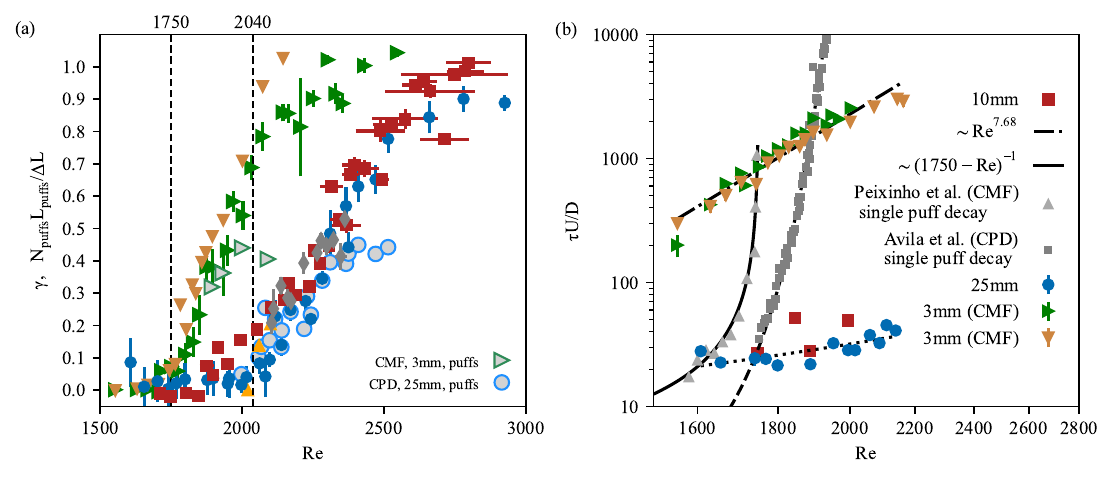}
\vspace{-0.5em}
\caption{(a) Plot of the steady-state turbulent fraction $\gamma$ for CMF and CPD pipe flows determined from pressure measurements and estimated from measuring the density of puffs. Both estimates agree until $\gamma \sim 0.5$ where in CPD flow the puffs are giving way to slugs. The difference between CPD and CMF flow can thus be attributed to the difference in puff density. (b) Log-log plot of the normalized decay times $\tau U/D$ vs. $Re$ for CMF and CPD flows. $\tau U/D$ appears to grow as a power-law with $Re$ for the CMF flows ($- \cdot - \cdot$), 
while for the CPD flows the slight increase in slope with $Re$ suggests an exponential fit ($\cdot\cdot\cdot$). The $L/D=4162$ CMF measurements were made upstream of the $L/D=4502$ measurements to investigate the effect of pipe length for CMF flow, which is imperceptible for the decay times and tends to increase $\langle \gamma \rangle$ by $\sim 5\%$ on average. The CMF decay times are larger than their CPD counterparts by an order of magnitude. It is curious that in the CMF pipe flow the interactions between puffs appears to enhance stability relative to the decay of single puffs, at least for $Re \lesssim 1750$, while the opposite is the case for CPD pipe flow. The most striking conclusion from both comparisons is that both the temporal and stationary behavior of CMF flow is qualitatively different from CPD flow.}
\label{fig:finalComparison}
\vspace{0.0em}
\end{figure}

Now we carry out a comparison between the CPD and CMF experimental results. With the CPD and CMF experiments combined we performed $\gtrsim$ 400 rehearsals over a period of $\sim$ 40 months in an air-conditioned laboratory environment, with the experimental run time collectively reaching $\sim$ 3.5$\times 10^5$ advective time units. 
The time series of the turbulent fraction $\gamma(t)$ for CPD (Fig.~\ref{fig:CPDTimeSeriesDecayTimes}(a)) and CMF (Fig.~\ref{fig:timeSeriesAndDecayAndMoodyCMF}(a)) shows behavior that is both similar but also strikingly different. Each individual curve begins with a large turbulent fraction, and then decays to a final state with fluctuations about a constant value. However, the numerical value of the decay times and $f$ (or $\gamma$) in the final state differ dramatically at the same $Re$. More precisely, $f$ and $\gamma$ are much higher for the CMF experiments than for the CPD experiments at comparable values of $Re$, which as we shall see has important consequences for the Moody diagram. Likewise the decay time is an order of magnitude larger in the CMF experiments. Understanding the differences between the quenched flow in the CMF and CPD experiments will thus require us to at least characterize the final state (the value of $f$ we use in the Moody diagram) and the initial decay.

The stationary behavior of the quenched CMF and CPD flows differs considerably. As shown in Fig.~\ref{fig:finalComparison}(a) The CMF turbulent fraction curve peels away from zero at a critical $Re_C \simeq 1750$ and continues to rise while the CPD curve remains near zero until $Re_C \simeq 2000$, thus yielding $Re_C$ that are close to previously reported values determined from experiments on individual puffs~\cite{peixinho2006decay,avila2011onset}. These differences are outside the statistical uncertainty and cannot be explained by differences in $L/D$ as the CPD span over an order of magnitude in $L/D$ and yet coincide, and artificially decreasing $L/D$ for the CMF flow by measuring the pressure further upstream also makes little difference for either $\tau U/D$ or $\langle \gamma \rangle$. Moreover the non-dimensional quenching time $\Delta T U/D$ for the CMF experiment is intermediate to the two CPD experiments, providing evidence that important features of the quench protocol do not influence the final results. We also recall again that the single curve traced out by our CPD quench experiments coincides with a maximal friction curve and with experiments that attempted to artificially create an arbitrarily long pipe, underscoring its robustness. Likewise the CMF quench curve in Fig.~\ref{fig:timeSeriesAndDecayAndMoodyCMF}(c) is distinct from all other data shown in Fig.~\ref{fig:moodyMayhem}, displayed in the background. We conclude that the origin of this difference is the driving mechanism for each flow.

Next we investigate the exponential decay times $\tau U/D$ vs. $Re$ in Fig.~\ref{fig:finalComparison}(b), where $\tau$ is determined by an exponential fit to $\gamma(t)$ until it decays halfway to the final state value. 
The quench decay times for CMF flow are significantly longer than for CPD flow, just as single puffs survive longer in CMF flow than in CPD flow at the same $Re$~\cite{peixinho2006decay,avila2011onset,kuik2010quantitative}. We note that the longer lifetimes of CMF flow may offer an opportunity to study the decay of turbulence at very low $Re$ and seek for special solutions to the Navier-Stokes equations such as the periodic states found in plane Poiseuille flow~\cite{reynolds1967finite,pekeris1967stability,herbert1979periodic}.

The coincidence between the $Re_C$ observed here and the values determined by examining individual puffs suggests that the dynamics of single puffs controls the critical $Re_C$~\cite{mukund2018critical}. As $Re$ increases beyond $Re_C$, however, it is likely that additional physics such as puff interactions influence the flow behavior. Similarly it is likely that the same interactions are the cause of the small quench decay rate relative to the single puff decay rate in CPD~\cite{avila2011onset,kuik2010quantitative}. To provide additional evidence that the differences observed here may be attributed to variations in puff dynamics, we analyzed the puff distributions in our CMF quench experiments.

We probed the puff distribution in the CMF pipe experiments using a pressure sensor which measured the pressure difference over a very small distance ($\sim6.66D$). As in Ref.~\cite{avila2011onset}, this allowed us to identify puffs as local spikes in the pressure (see App.~\ref{sec:puffsCMF}). Comparing velocity measurements of the axial velocity profile of puffs from different literature sources~\cite{darbyshire1995transition,hof2010eliminating,nishi2008laminar} as well as our own measurements of puffs in CPD flows, we determined that their spatial extent does not differ within experimental accuracy (see $SM$). Thus using the size of the turbulent core of the puff $L_{\rm{puffs}} \simeq 16.6D$, which is the length that contributes to the turbulent friction~\cite{cerbus2018laws}, we estimate that the turbulent fraction can be independently estimated from the number of puffs as $N_{\rm{puffs}} L_{\rm{puffs}} / \Delta L$, where $\Delta L$ is the distance over which the measurement is made. With this estimate we find in Fig.~\ref{fig:finalComparison}(a) excellent agreement between $N_{\rm{puffs}} L_{\rm{puffs}} / \Delta L$ and curves of average turbulent fraction $\gamma$ for both the CMF and CPD, estimated from pressure measurements. The agreement is maintained up until $\gamma \simeq 0.5$, which for the CPD experiments corresponds to the range of $Re$ where puffs transition into slugs.

The data in Fig.~\ref{fig:finalComparison}(a) shows that, at least for a range of $Re$, the origin of the different $\gamma$ and thus $f$ values for the CMF and CPD curves may possibly be attributed to the different number of puffs ($N_{\rm{puffs}}$) that survive the quench. However, we showed that in our CPD experiments the density of puffs is maximal after the quench, and yet the value of $f$ for CPD is generally lower than for CMF at the same $Re$. A possible resolution to this apparent contradiction is that CPD and CMF flows have different maximal puff densities. Since the size of the puffs does not appear to differ significantly between CPD and CMF flows (see $SM$), this corresponds to a smaller average spacing between puffs in CMF flow. We submit that the different decay time scales and permissible density of puffs in CMF flow combine to produce the different curves shown in Figs.~\ref{fig:finalComparison}. Compared to CPD flow, puffs have been examined comparatively less in CMF flow, and we propose that further investigation may shed more light on these differences.

\section{Summary and Conclusions}
\label{sec:conclusion}

\begin{figure}
\centering
\vspace{-0em}
\includegraphics[width=0.5\linewidth]{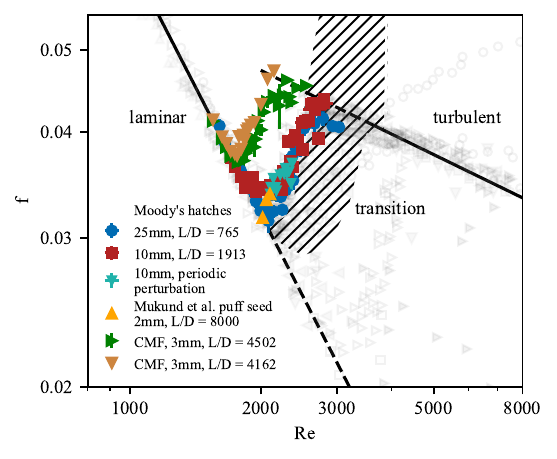}
\vspace{-0.5em}
\caption{Plot of $f$ vs. $Re$ for both experimental driving protocols. Vertical error bars are the standard deviation of the mean, and horizontal error bars are the standard deviation of the instantaneous fluctuations in $Re$ after the quench. The respective deviations from the laminar curve occur at $Re$ close to critical values previously found in studies of single puff decay ($Re_c^{\rm{CMF}} \simeq 1750$, $--$; $Re_c^{\rm{CPD}} \simeq 2040$, $\cdot \cdot \cdot$)~\cite{peixinho2006decay,avila2011onset}. Here we conclude that two distinct curves serve to fill the gap in the Moody diagram, corresponding to the two methods of driving the flow.}
\label{fig:finalMoody}
\vspace{0.0em}
\end{figure}

In conclusion, we have performed a large number of experiments which have helped uncover consistent behaviour in the gap in the Moody diagram. When the flow is driven using a constant pressure difference, as is commonly the case in many laboratory experiments and nearly always in industrial settings, we find a single curve that corresponds to the long-time and maximal value of the turbulent fraction in this region. This finding serves as a starting point for future investigations of friction or turbulent fraction in the transition to turbulence in CPD pipe flow, and could also be used as a diagnostic for industrial systems. On the other hand, we also found that when the flow is driven using a constant mass flux, a different curve emerges, which sheds light on the rich and complex nature of pipe flow in the transitional regime. Seeking a single, universal curve, we instead uncovered evidence for two distinct curves, as shown in Fig.~\ref{fig:finalMoody}. Given the historical and practical background of indeterminancy that led Moody to use hatches to cover the transitional range of $Re$. Establishing such systematic behavior in the transitional regime provides an advance which will aid understanding the sources of turbulence in pipe flows.

\section{Acknowledgements}

We thank Jorge Peixinho, Pinaki Chakraborty, and Hamid Kellay for helpful comments. R.T.C. gratefully acknowledges funding from the Horizon 2020 program under the Marie Skłodowska-Curie Action Individual Fellowship (MSCAIF) No. 793507, the support of JSPS (KAKENHI Grant No. 17K14594), and the support of the Okinawa Institute of Science and Technology (OIST) where the experiments were carried out.

\appendix

\section{Sensitivity to quench protocol}
\label{sec:appA}

\begin{figure}
\vspace{-0em}
\centering
  \includegraphics[width=0.8\textwidth]{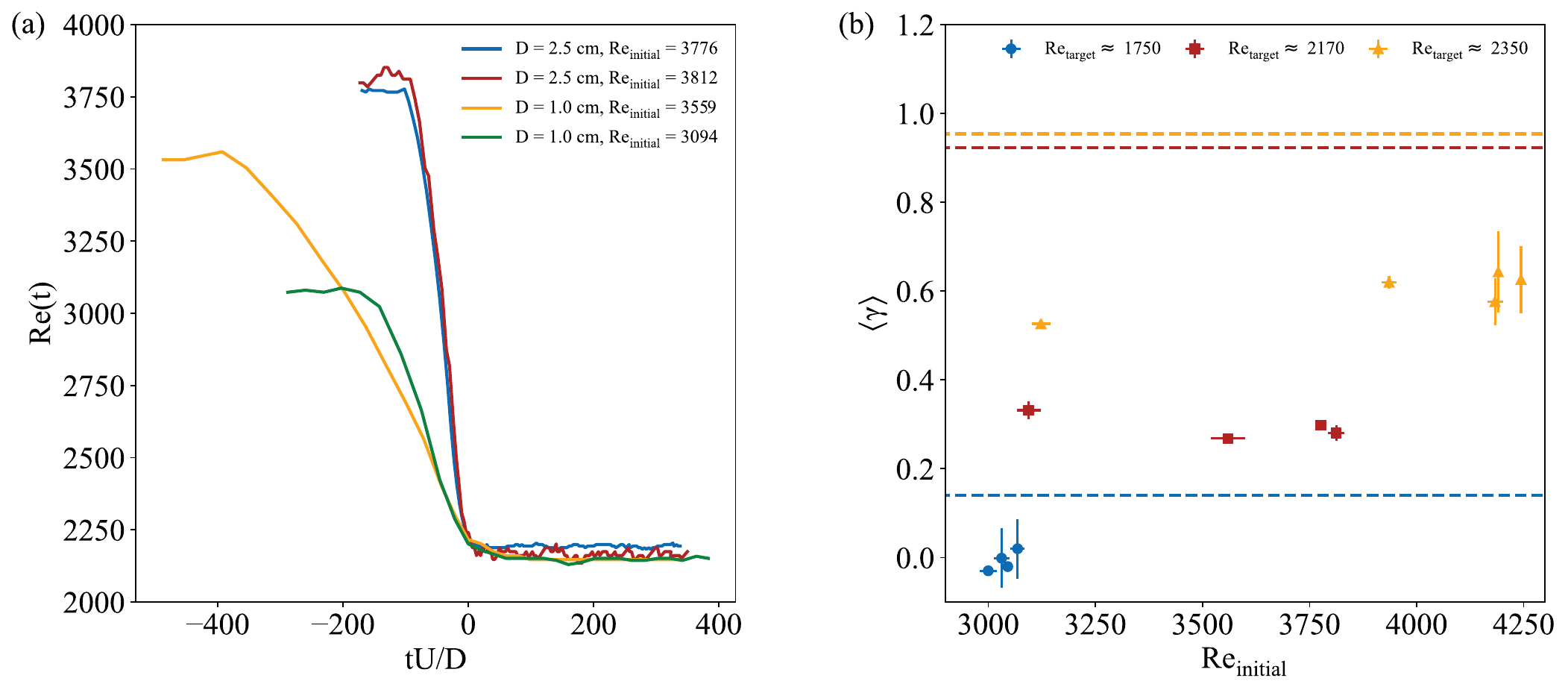}
\caption{(a) Plot of $Re(t)$ versus $tU/D$ for several CPD quench experiments, both $D$ = 2.5 cm and $D$ = 1 cm with the same target quench value of $Re \simeq 2170$. Although the initial $Re_{\rm{initial}}$ and the quench time $\Delta T U/D$ can differ significantly, the effect on the final quenched state, $\langle \gamma \rangle$ is not noticeable. (b) Plot of $\langle \gamma \rangle$ versus $Re_{\rm{initial}}$ for several quench target $Re_{\rm{target}}$ which includes both $D$ = 1 cm and $D$ = 2.5 cm. The variation is negligible, in particular compared with the value of $\langle \gamma \rangle$ for the same $Re_{\rm{target}}$ for the CMF flow (dashed lines, $--$, of the same color).}
\vspace{-1em}
\label{fig:quenchSensitivity}
\end{figure}

Although our initial condition is turbulent flow in all experiments, there nevertheless exist differences in the quench protocol as described in the main text and in Secs. I, II. Two significant differences are the time taken to quench $\Delta T U / D$ and the initial $Re$ of the turbulent flow before quenching, both of which vary between the three sets of experiments (2 CPD and 1 CMF). Because the quench times of all three are (at $Re = 2000$): $\Delta T U/ D \simeq 33$ ($D$ = 2.5 cm CPD), $\Delta T U/ D \simeq 205$ ($D$ = 1 cm CPD), and $\Delta T U/ D \simeq 82$ ($D$ = 0.3 cm CMF). Thus we argue that the largest difference occurs between the two CPD flows, which nevertheless show broad agreement in the results for both $\tau U / D$ and $\langle \gamma \rangle$, shown in Fig.~\ref{fig:finalComparison}. In Fig.~\ref{fig:quenchSensitivity} we test both the sensitivity to $\Delta T U / D$ as well as the initial turbulent $Re$, $Re_{\rm{initial}}$. Fig.~\ref{fig:quenchSensitivity}a shows several time series with the same target $Re$, $Re_{\rm{target}} \simeq 2170$, but different $\Delta T U / D$ and $Re_{\rm{initial}}$. Also shown in Fig.~\ref{fig:quenchSensitivity}b for this $Re_{\rm{target}}$ and several others, $\langle \gamma \rangle$ changes insignificantly, especially in comparison with the value for the CMF experiments at the same $Re_{\rm{target}}$ (dashed lines). The results shown in Fig.~\ref{fig:quenchSensitivity} demonstrate that our quench results are insensitive to the details of the quench protocol.

\section{Ancillary experiments with periodic perturbations}

\begin{figure}
\vspace{-0em}
\centering
  \includegraphics[width=0.4\textwidth]{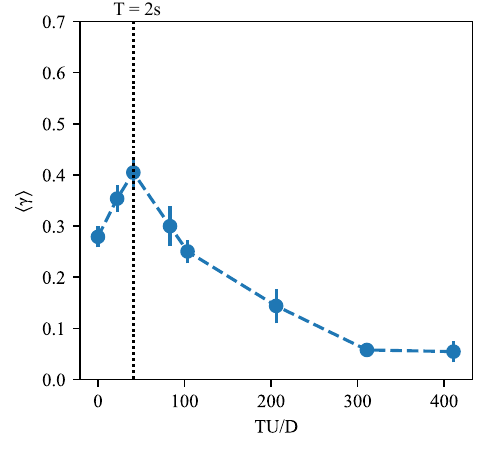}
\caption{The average friction factor $\langle f \rangle$ for the train of puffs as a function of forcing period $T U /D$. Here $Re \sim 2231$. The maximum $\langle f \rangle$ (which will yield the maximum $\langle \gamma \rangle$ was found for a period of $T = 2s$, which we thus used for all $Re$.}
\label{fig:periodicPerturbation}
\end{figure}

Here we describe the ancillary experiments used to determine a putative maximum $\langle \gamma \rangle$ curve in Fig. 3b in the main text. Using the $D = 1$ cm CPD pipe setup, we connected two low-pressure syringe pumps (New Era) to the pressure tap furthest upstream, 4m ($\simeq 404D$) away from the entrance of the pipe and 10m ($\simeq 1010D$) upstream from the same pressure measurement section used for the quenching experiments. The pumps are synchronized so that one injects while the other simultaneously withdraws the same amount of fluid at the same rate. The maximum frequency of the syringe pump was 1 Hz. Thus this syringe pump could not be used for a similar study with our CMF setup since the minimum puff spacing we could achieve is $T U /D \sim 200$ (at $Re = 2000$). After establishing a laminar flow in the pipe and calibrating the pressure sensors, we created a train of puffs by injecting and withdrawing a small amount of fluid (0.15 mL) at a rate of 30mL/s, similar to Refs.~\cite{cerbus2018laws,samanta2011experimental}. The subsequent friction factor $f$ is plotted as a function of spacing $T U /D$ in Fig.~\ref{fig:periodicPerturbation} at $Re \sim 2231$. The maximum value of $\langle f \rangle$ (and thus $\langle \gamma \rangle$) is found for $T U /D \simeq 40$ or $T = 2$s. We used this same $T$ for each $Re$ for the results in Fig. 3b in the main text. As found in Ref.~\cite{samanta2011experimental}, the maximum value of $f$ (and thus our $\langle \gamma \rangle$) is not sensitive to changes in the period $T$ in the neighborhood of the maximum.

\section{Identification of puffs in CMF}
\label{sec:puffsCMF}

\begin{figure}
\vspace{-0em}
\centering
  \includegraphics[width=0.9\textwidth]{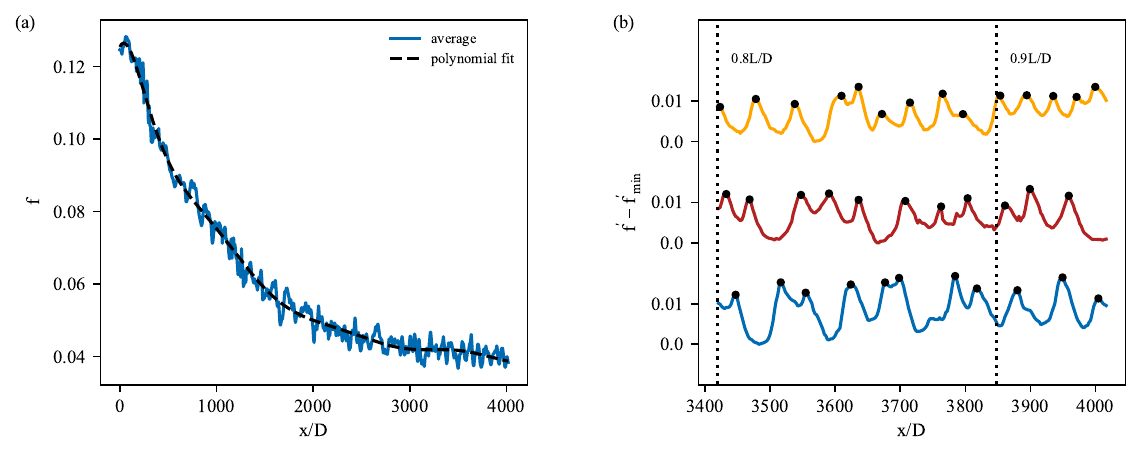}
\caption{(a) Example of a friction time series at $Re\simeq1894$ from pressure measurements made over a shorter distance. The ensemble-averaged curve has large fluctuations which we identify with puffs as in Ref.~\cite{avila2011onset} We use a polynomial fit to estimate the average behavior and then subtract this from the friction time series of individual runs. (b) Individual friction curves with the estimated mean subtracted off and shifted vertically for visualization. We identify the peaks with individual puffs.}
\label{fig:puffsCMF}
\end{figure}

In the CMF experiments we avoided performing LDV velocity measurements due to the difficulty in pipe maintenance and cleaning if tracer particles were added to the very small diameter pipe (3mm). Instead we probed small-scale features of the flow by using our custom-made junctions for the tubing sections. These are 3D-printed and machined with pressure taps a distance of 20 mm ($\simeq 6.7D$) apart to enable us to measure pressure fluctuations with a higher spatial resolution. A similar method was used to identify puffs in experiments on single puff lifetimes in a CPD pipe flow experiment~\cite{avila2011onset}. The pressure measurements across this short distance, which were performed at the same time as some of the experiments reported in the main text, reproduce the overall behavior of the measurements made over a larger span ($\gtrsim 300D$). However, these measurements yield much larger fluctuations. In Fig.~\ref{fig:puffsCMF}(a) we show the ensemble average of three sequential quench experiments at $Re\simeq1894$. We make a polynomial fit to the averaged curve, and then subtract this from each instantaneous $f(x/D)$ curve (equivalently $f(tU/D)$ curve), yielding a fluctuation curve $f'$. We plot these three curves in Fig.~\ref{fig:puffsCMF}(b) where we have set the minimum value to be zero and shifted them vertically for visualization. We then identify the peaks of this curve as the individual puffs, similar to Ref.~\cite{avila2011onset}. This gives us the puff number density used to estimate the pressure contribution of puffs in Fig.~\ref{fig:finalComparison}(a).

\bibliography{quenchingPipeFlow}

\end{document}